\documentclass[aps,pra,preprint,groupedaddress,showpacs,showkeys]{revtex4}
\usepackage{amssymb,bm}
\usepackage{graphicx}
\usepackage{amsmath}
\allowdisplaybreaks

\begin{document}

\title{Charge asymmetry in the differential cross section of high-energy
 $e^+e^-$ photoproduction  in the  field of a heavy atom}

\author{R.N. Lee}\email{R.N.Lee@inp.nsk.su}
\author{A. I. Milstein}\email{A.I.Milstein@inp.nsk.su}
\author{V.M. Strakhovenko}\email{V.M.Strakhovenko@inp.nsk.su}

\affiliation{Budker Institute of Nuclear Physics, 630090 Novosibirsk, Russia}

\date{\today}

\begin{abstract}
First quasiclassical correction to the differential cross section of  high-energy  electron-positron photoproduction  in the electric field of a heavy atom is obtained with the exact account of the field. This correction is responsible for the charge asymmetry ${\cal A}$ in this process. When the transverse momentum of at least one of the produced particles is much larger than the electron mass $m$, the charge asymmetry can be as large as tens percent.  We also estimate the contribution $\tilde{\cal A}$ to the charge asymmetry coming from the Compton-type diagram. For heavy nuclei, this contribution is negligible. For light nuclei, $\tilde{\cal A}$ is noticeable only when the angle between the momenta of electron and positron is of order of $m/\omega$ ($\omega$ is the photon energy) while the transverse momenta of both particles are much larger than $m$.
\end{abstract}

\pacs{32.80.-t, 12.20.Ds}

\keywords{ $e^+e^-$ photoproduction, Coulomb corrections, charge asymmetry}

\maketitle



\section{Introduction}
The production of an electron-positron  pair by a photon in an atomic field is one of the most important processes  of QED. Because of its importance  for various applications, see Refs.~\cite{HGO1980,Hubbell2000}, this process has been investigated in  numerous  theoretical and experimental papers.  The cross section of the process in the Born approximation is known for arbitrary energy $\omega$ of the incoming photon, Refs.~\cite{BH1934,Racah1934} (we set $\hbar=c=1$ throughout the paper).  For heavy atoms, it is  necessary to take into account the  higher-order terms of the perturbation theory with respect to the parameter $Z\alpha$ (Coulomb corrections), where $Z$ is the atomic charge number, $\alpha=e^2\approx 1/137$ is the fine-structure constant, and $e$ is the electron charge. The formal expression for the Coulomb corrections, exact in $Z\alpha$ and $\omega$, was derived in Ref.~\cite{Overbo1968}. However, the numerical computations based on this expression  becomes more and more difficult when $\omega$ is increasing, and the numerical results have been obtained so far only for $\omega<12.5\,$ MeV, Ref. \cite{SudSharma2006}.

Fortunately, in the high-energy region $\omega\gg m$ ($m$ is the electron mass), a completely different approach, which greatly simplifies  the calculations,  can be used. As a result of this approach, a simple expression for the Coulomb corrections was obtained in Refs.\cite{BM1954,DBM1954} in the leading approximation with respect to $m/\omega$. However, this result has good accuracy only at energies $\omega \gtrsim 100\,$MeV. For a long time, the  description of the Coulomb corrections for the total cross section at intermediate photon energies ($5 \div 100\,$ MeV) was based  on the expression obtained in Ref.~\cite{Overbo1977}. This expression is actually an extrapolation of the results obtained at $\omega<5\,$MeV. Recently,  the first corrections of the order of $m/\omega$ to the spectrum, as well as to the total cross section, of $e^+e^-$ photoproduction in a strong atomic field were derived  in Ref.~\cite{LMS2004}. The correction to the spectrum was obtained in the region where both produced particles are relativistic. It turns out that this correction is antisymmetric with respect to replacement $\varepsilon_+\leftrightarrow\varepsilon_-$, where $\varepsilon_+$ and $\varepsilon_-$ are the energy of the positron and the electron, respectively, so that the correction to the total cross section  comes from the region close to the end of spectrum where  $\varepsilon_+\sim m$ or $\varepsilon_-\sim m$. In Ref. \cite{LMS2004}, the correction to the total cross section was obtained by means of the dispersion relation for the forward Delbr\"uck scattering amplitude. The account for this correction leads to agreement between the theoretical prediction and available experimental data at intermediate photon energies, Ref.~\cite{LMS2003}.
The electron (positron) spectrum in the process of $e^+e^-$ photoproduction in a strong Coulomb field in the case $\varepsilon_+\sim m$ or $\varepsilon_-\sim m$ and $\omega\gg m$ was investigated in Ref.~\cite{DM10}. It was shown  that the Coulomb corrections drastically differ from those obtained in the region where $\varepsilon_+\gg m$ and $\varepsilon_-\gg m$. Integration of the spectrum in Ref.~\cite{DM10} has confirmed the result for the correction to the total cross section obtained in Ref.~\cite{LMS2003} by means of the dispersion relation.

In the present paper, we calculate, exactly in $\eta=Z\alpha$,  the next-to-leading correction with respect to $m/\omega$ to the differential cross section of   electron-positron  pair production by a high-energy photon in a strong atomic field. The correction, being the odd function of $\eta$, gives rise to the charge asymmetry in this process. The leading term of the cross section  obtained in Refs.~\cite{BM1954,DBM1954} is the even function of  $Z\alpha$ and  does not contribute to the charge asymmetry. For the photon energy below the threshold of $\pi$-meson photoproduction, we also estimate the contribution of the Compton-type amplitude  to the $e^+e^-$ photoproduction cross section. The corresponding term contributes to the charge asymmetry as well.

\section{General discussion}

\begin{figure}[h]
\includegraphics[width=5cm]{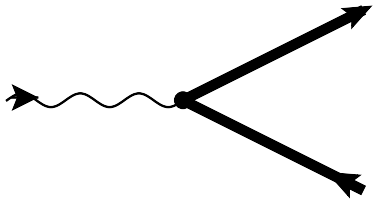}
\caption{Diagram for pair production by a photon in a strong Coulomb field. Thick lines correspond to the positive- and negative-energy solutions of the Dirac equation in the Coulomb field.}\label{fig:dia1}
\end{figure}

The cross section of $e^+e^-$ pair production by a photon in an
external field reads (see, e.g., Ref.~\cite{BLP82} )
\begin{equation}\label{eq:cs}
d\sigma=\frac{\alpha}{(2\pi)^4\omega}\,d\bm{p}\,d\bm{q}\,\delta
(\omega - \varepsilon_p -\varepsilon_q)|M_{\lambda_1\lambda_2}|^{2}\,,
\end{equation}
where $\varepsilon_{ p}=\sqrt{ p^2+m^2}$,
 $\bm p$ and $\bm q$ are the electron and
positron momenta, respectively. The matrix element $M_{\lambda_1\lambda_2}$, corresponding to the diagram shown in Fig.\ref{fig:dia1} has the form
\begin{equation}\label{M12}
M_{\lambda_1\lambda_2}\,=\,\int d\bm r \,\bar u_{\lambda_1\bm p }^{(out)}(\bm r )\,\bm\gamma\cdot
\bm e\,v _{\lambda_2\bm q}^{(in)}(\bm r )\exp{(i\bm k\cdot\bm r )}\,\,.
\end{equation}
Here $ u_{\lambda_1\bm p}^{(out)}(\bm r )$ is a positive-energy  solution and $v_{\lambda_2\bm q}^{(in)}(\bm r )$ is a
negative-energy solution of the Dirac equation in the external field, $\lambda_1=\pm 1$ and $\lambda_2=\pm 1$ enumerate the independent solutions of the Dirac equation,
$\bm e$ is the photon polarization vector,   $\bm k$ is the photon
momentum,  $\gamma^\mu$ are the Dirac matrices.
We remind that the asymptotic forms of $ u_{\lambda\bm p}^{(out)}(\bm r )$ and $ v_{\lambda\bm p}^{(out)}(\bm r )$ at large $\bm r$ contain the plane waves and the spherical convergent waves, while the   asymptotic forms of $ u_{\lambda\bm p}^{(in)}(\bm r )$ and $ v_{\lambda\bm q}^{(in)}(\bm r )$ at large $\bm r$ contain the plane waves and the spherical divergent waves.
In order to obtain the wave functions in the leading quasiclassical approximation and the first quasiclassical
correction to it, we exploit  the  convenient integral representation for the exact wave functions in the Coulomb field suggested in Ref. \cite{MT2004}. The derivation of this representation is based on the relations
\begin{eqnarray}\label{Green1}
&&\lim_{ r_1\to \infty }G({\bf r}_2,{\bf r}_1|\varepsilon_{ p})=
-\frac{\exp{(ipr_1+i\eta_p\ln(2pr_1))}}{4\pi r_1}
\sum_{\lambda=1,2} 2 \varepsilon_{ p}u_{\lambda{\bf p}}^{(in)}({\bf
r}_2)\bar{u}_{\lambda{\bf p}}\,,\quad \bm p=-p\bm n_1\,
,\nonumber\\
&&\lim_{ r_1\to \infty }G({\bf r}_2,{\bf r}_1|-\varepsilon_{ p})=
\frac{\exp{(ipr_1-i\eta_p\ln(2pr_1))}}{4\pi r_1}
\sum_{\lambda=1,2}2 \varepsilon_{ p}v_{\lambda{\bf p}}^{(in)}({\bf
r}_2)\bar{v}_{\lambda{\bf p}}\,,\quad \bm p=p\bm n_1\,
,\nonumber\\
&& u_{\lambda{\bf p}}=\sqrt{\frac{\varepsilon_p+m}{2\varepsilon_p}}
\begin{pmatrix}
\phi_\lambda\\
\dfrac{{\boldsymbol \sigma}\cdot {\boldsymbol
p}}{\varepsilon_p+m}\phi_\lambda\end{pmatrix}\, ,\quad
v_{\lambda{\bf p}}=\sqrt{\frac{\varepsilon_p+m}{2\varepsilon_p}}
\begin{pmatrix}
\dfrac{{\boldsymbol \sigma}\cdot {\boldsymbol
p}}{\varepsilon_p+m}\chi_\lambda\\
\chi_\lambda\end{pmatrix}\, ,\quad \eta_p=Z\alpha\frac{\varepsilon_{ p}}{p}\,,
\end{eqnarray}
and also
\begin{eqnarray}\label{Green2}
&&\lim_{ r_2\to \infty }G({\bf r}_2,{\bf r}_1|\varepsilon_{ p})=
-\frac{\exp{(ipr_2+i\eta_p\ln(2pr_2))}}{4\pi r_2}
\sum_{\lambda=1,2}2 \varepsilon_{ p}u_{\lambda{\bf p}}\bar{u}_{\lambda{\bf p}}^{(out)}({\bf
r}_1)\,,\quad \bm p=p\bm n_2\,
,\nonumber\\
&&\lim_{ r_2\to \infty }G({\bf r}_2,{\bf r}_1|-\varepsilon_{ p})=
\frac{\exp{(ipr_2-i\eta_p\ln(2pr_2))}}{4\pi r_2}
\sum_{\lambda=1,2}2 \varepsilon_{ p} v_{\lambda{\bf p}}\bar{v}_{\lambda{\bf p}}^{(out)}({\bf
r}_2)\,\,,\quad \bm p=-p\bm n_2\,,
\end{eqnarray}
where $G({\bf r}_2,{\bf r}_1|\varepsilon)$ is the Green function  of the Dirac
  equation in the Coulomb field, $\bm n_1=\bm r_1/r_1$ and $\bm n_2=\bm r_2/r_2$.
A convenient integral representation for $G({\bf r}_2,{\bf r}_1|\varepsilon)$ was obtained in Ref.~\cite{MS82}. Using Eqs. (19)-(22) of that paper, we arrive at the following result for the wave  functions $u_{\lambda{\bf p}}^{(in)}({\bf r})$ and
$v_{\lambda{\bf p}}^{(in)}({\bf r})$:
\begin{eqnarray}\label{wf1}
&&u_{\lambda{\bf p}}^{(in)}({\bf r})=\frac{\exp(ipr)}{pr}
\int\limits_0^\infty
dt\,t^{-2i\eta_p-1}{\mbox e}^{it^2}\nonumber\\
&&\times\left[S_B(-x_{\bm p},\,pr)\left(t^2-Z\alpha\frac{m}{p}\gamma^0\right)
(1-R_1)+iS_A(-x_{\bm p},\,pr)(1+R_1)\right]u_{\lambda{\bf p}}\,,\nonumber\\
&&v_{\lambda{\bf p}}^{(in)}({\bf r})=\frac{\exp(ipr)}{pr}
\int\limits_0^\infty
dt\,t^{2i\eta_p-1}{\mbox e}^{it^2}\nonumber\\
&&\times\left[S_B(x_{\bm p},\,pr)\left(t^2-Z\alpha\frac{m}{p}\gamma^0\right)
(1-R_2)+iS_A(x_{\bm p},\,pr)(1+R_2)\right]v_{\lambda{\bf p}}\,,\nonumber\\
&&S_A(x,\,\rho)=\sum_{l=1}^{\infty}{\mbox e}^{-i\pi\nu}J_{2\nu}(2t\sqrt{2\rho})\,l\frac{d}{dx}(P_l(x)+P_{l-1}(x))\, ,\nonumber\\
&&S_B(x,\,\rho)=\sum_{l=1}^{\infty}{\mbox e}^{-i\pi\nu}J_{2\nu}(2t\sqrt{2\rho})\frac{d}{dx}(P_l(x)-P_{l-1}(x))\quad ,\nonumber\\
&& x_{\bm p}=\frac{\bm r\cdot \bm p}{rp}\, ,\quad
R_{1,2}=(m\pm\gamma^0\varepsilon_p)\frac{\bm \gamma\cdot\bm r}{pr}
\,.
\end{eqnarray}
Here $P_l(x)$ is the Legendre polynomial, $J_{2\nu}$ is the Bessel
function, $\nu=\sqrt{l^2-(Z\alpha)^2}$.  For the wave  functions $\bar u_{\lambda{\bf p}}^{(out)}({\bf r})$ and
$\bar v_{\lambda{\bf p}}^{(out )}({\bf r})$ we obtain:
\begin{eqnarray}\label{wf2}
&&\bar u_{\lambda{\bf p}}^{(out)}({\bf r})=\frac{\exp(ipr)}{pr}
\int\limits_0^\infty
dt\,t^{-2i\eta_p-1}{\mbox e}^{it^2}\nonumber\\
&&\times\bar u_{\lambda{\bf p}}\left[S_B(x_{\bm p},\,pr)
(1+R_2)\left(t^2-Z\alpha\frac{m}{p}\gamma^0\right)+iS_A(x_{\bm p},\,pr)(1-R_2)\right]\,,\nonumber\\
&&\bar v_{\lambda{\bf p}}^{(out)}({\bf r})=\frac{\exp(ipr)}{pr}
\int\limits_0^\infty
dt\,t^{2i\eta_p-1}{\mbox e}^{it^2}\nonumber\\
&&\times\bar v_{\lambda{\bf p}}\left[S_B(-x_{\bm p},\,pr)
(1+R_1)\left(t^2-Z\alpha\frac{m}{p}\gamma^0\right)+iS_A(-x_{\bm p},\,pr)(1-R_1)\right]\,.
\end{eqnarray}
The integrals over the variable $t$ in Eqs. (\ref{wf1}) and (\ref{wf2}) can be expressed via the confluent
hypergeometric functions. However, the forms   (\ref{wf1}) and (\ref{wf2}) of the wave functions are  more convenient for applications than
the conventional ones. The results (\ref{wf1}) and (\ref{wf2}) are in agreement with the well-known solutions of the Dirac equation in the Coulomb
field.

\section{Calculation of the matrix element}

Let us introduce the functions
\begin{eqnarray}\label{fun123}
&&F_A(\bm r,\,\bm p,\,\eta)=i\frac{\exp(ipr)}{pr}
\int\limits_0^\infty dt\,t^{-2i\eta-1}{\mbox e}^{it^2}S_A(x_{\bm p},\,pr)\,,\nonumber\\
&&F_B(\bm r,\,\bm p,\,\eta)=\frac{\exp(ipr)}{pr}
\int\limits_0^\infty dt\,t^{-2i\eta-1}{\mbox e}^{it^2}S_B(x_{\bm p},\,pr)\,,\nonumber\\
&&{\tilde F}_B(\bm r,\,\bm p,\,\eta)=\frac{\exp(ipr)}{pr}
\int\limits_0^\infty dt\,t^{-2i\eta+1}{\mbox e}^{it^2}S_B(x_{\bm p},\,pr)\,.
\end{eqnarray}

In terms of the functions (\ref{fun123}), the wave functions $v_{\lambda{\bf p}}^{(in)}({\bf r})$
 and $\bar u_{\lambda{\bf p}}^{(out)}({\bf r})$ have the form
\begin{eqnarray}\label{vu2}
&&\bar u_{\lambda{\bf p}}^{(out)}({\bf r})=\Bigg(\phi^+{\cal R}_1^{(+)}\,,\,
-\phi_\lambda^+{\cal R}_2^{(+)}\Bigg)\,,\quad
v_{\lambda{\bf q}}^{(in)}({\bf r})=
\begin{pmatrix}
{\cal R}_2^{(-)}\chi_\lambda\\
{\cal R}_1^{(-)}\chi_\lambda\end{pmatrix}\, ,\nonumber\\
&&{\cal R}_1^{(+)}=\sqrt{\frac{\varepsilon_p+m}{2\varepsilon_p}}\left[(\tilde F_B^{(+)}-\frac{Z\alpha m}{p} F_B^{(+)})(1+\bm\sigma\cdot\hat{\bm p}\,\bm\sigma\cdot\bm n)
+F_A^{(+)}(1-\bm\sigma\cdot\hat{\bm p}\,\bm\sigma\cdot\bm n)\right]\,,\nonumber\\
&&{\cal R}_2^{(+)}=\sqrt{\frac{\varepsilon_p-m}{2\varepsilon_p}}\left[(\tilde F_B^{(+)}+\frac{Z\alpha m}{p} F_B^{(+)})(\bm\sigma\cdot\hat{\bm p}+\bm\sigma\cdot\bm n)+F_A^{(+)}(\bm\sigma\cdot\hat{\bm p} -\bm\sigma\cdot\bm n)\right]\,,\nonumber\\
&&{\cal R}_1^{(-)}=\sqrt{\frac{\varepsilon_q+m}{2\varepsilon_q}}\left[(\tilde F_B^{(-)}+\frac{Z\alpha m}{q} F_B^{(-)})(1+\bm\sigma\cdot\bm n\,\bm\sigma\cdot\hat{\bm q})
+F_A^{(-)}(1-\bm\sigma\cdot\bm n\,\bm\sigma\cdot\hat{\bm q})\right]\,,\nonumber\\
&&{\cal R}_2^{(-)}=\sqrt{\frac{\varepsilon_q-m}{2\varepsilon_q}}\left[(\tilde F_B^{(-)}-\frac{Z\alpha m}{q} F_B^{(-)})(\bm\sigma\cdot\hat{\bm q}+\bm\sigma\cdot\bm n)+F_A^{(-)}(\bm\sigma\cdot\hat{\bm q} -\bm\sigma\cdot\bm n)\right]\,,
\end{eqnarray}
where $\bm n=\bm r/r$, $\hat{\bm p}={\bm p}/p$, $\hat{\bm q}={\bm q}/q$        and
\begin{eqnarray}
&&F_A^{(+)}= F_A(\bm r,\,\bm p,\,\eta_p)\,,\quad F_B^{(+)}= F_B(\bm r,\,\bm p,\,\eta_p)\,,\quad\tilde F_B^{(+)}=\tilde F_B(\bm r,\,\bm p,\,\eta_p)\,,\nonumber\\
&&F_A^{(-)}= F_A(\bm r,\,\bm q,\,-\eta_q)\,,\quad F_B^{(-)}= F_B(\bm r,\,\bm q,\,-\eta_q)\,,\quad\tilde F_B^{(-)}=\tilde F_B(\bm r,\,\bm q,\,-\eta_q)\,.
\end{eqnarray}
Then the matrix element   $M_{\lambda_1\lambda_2}$, Eq. (\ref{M12}), is
\begin{eqnarray}\label{M12N}
M_{\lambda_1\lambda_2}&=&\int d\bm r \,\exp{(i\bm k\cdot\bm r )}\phi_{\lambda_1}^{+}
\left[{\cal R}_1^{(+)}\,\bm\sigma\cdot\bm e\, {\cal R}_1^{(-)}
+{\cal R}_2^{(+)}\,\bm\sigma\cdot\bm e\, {\cal R}_2^{(-)}\right]\chi_{\lambda_2}\,.
\end{eqnarray}
For any vector $\bm X$ we introduce the notation  $\bm X_\perp=\bm X-(\bm\nu\cdot\bm X)\bm\nu$, $\bm\nu=\bm k/k$, and write the matrix element $M_{\lambda_1\lambda_2}$ in the form
\begin{equation}\label{M12new}
M_{\lambda_1\lambda_2}=\phi_{\lambda_1}^{+}[(a_0+a_1)+\bm\sigma\cdot(\bm b_0+\bm b_1)]\chi_{\lambda_2} \,,
\end{equation}
where $a_0$ and $\bm b_0$ are linear in  $\bm\theta_p={\hat{\bm p}}_\perp$, $\bm\theta_q={\hat{\bm q}}_\perp$,
 $m/\varepsilon_p$ and $m/\varepsilon_q$ , while
 $a_1$ and $\bm b_1$ are quadratic in these variables. We have
\begin{eqnarray}\label{aabb}
&& a_0=2i[\bm\nu\times \bm e]\cdot\bm g^{(-)}\,,\quad \bm b_0=2\bm e\cdot \bm g^{(+)}\,\bm\nu+
2 g\,\bm e\,,\nonumber\\
&& a_1=i[\bm\nu\times \bm e]\cdot(\bm\theta_p-\bm\theta_q)\, g\,,\nonumber\\
&&\bm b_1=\bm e\cdot\bm g^{(+)}\,(\bm\theta_p+\bm\theta_q)+
\bm e\cdot(\bm\theta_p-\bm\theta_q)\,\bm g^{(-)}\nonumber\\
&&-(\bm\theta_p-\bm\theta_q)\cdot\bm g^{(-)}\,\bm e
-\bm e\cdot(\bm\theta_p+\bm\theta_q)\,g\bm\nu\,.
  \end{eqnarray}
Here
\begin{eqnarray}\label{funnew}
\bm g^{(\pm)}&=&\int  d\bm r\,\exp(i\bm k\cdot\bm r)F_A^{(+)}\left[(\bm n_{\perp}+\bm\theta_q){\tilde F}_B^{(-)}
-\bm n_{\perp}{F}_A^{(-)}\right]\nonumber\\
&&\pm\int  d\bm r\,\exp(i\bm k\cdot\bm r)F_A^{(-)}
\left[(\bm n_{\perp}+\bm\theta_p){\tilde F}_B^{(+)}
-\bm n_{\perp}{F}_A^{(+)}\right]\,,\nonumber\\
&&g=\frac{m\omega}{\varepsilon_p\varepsilon_q}\,\int  d\bm r\,\exp(i\bm k\cdot\bm r)F_A^{(+)} F_A^{(-)}\,.
\end{eqnarray}

In Ref. \cite{LMS2004}, the following expressions for the sum $S_A$ and $S_B$, which take into account the leading terms and first quasiclassical corrections, have been  obtained
\begin{eqnarray}\label{SASB}
&&S_A(x,\rho)=-\frac{y^2}{8}J_0\left(y\sqrt{\frac{1+x}{2}}\right)\left[1+\frac{i\pi (Z\alpha)^2}{y}\right]\,,\nonumber\\
&&S_B(x,\rho)=-\frac{y}{2\sqrt{2(1+x)}}J_1\left(y\sqrt{\frac{1+x}{2}}\right)\left[1+\frac{i\pi (Z\alpha)^2}{y}\right]\,,
  \end{eqnarray}
where $y=2t\sqrt{2\rho}$. These formulas are obtained for $y\gg 1$ and $1+x\ll 1$. Substituting Eq. (\ref{SASB}) to
Eq. (\ref{fun123}) and taking the integrals over the variable $t$ we find
\begin{eqnarray}\label{fun123QC}
&&F_A(\bm r,\,\bm p,\,\eta)=\frac{1}{2}\exp\left(\frac{\pi\eta}{2}-i\bm p\cdot\bm r\right)
[\Gamma(1-i\eta)F(i\eta,1,\,i(pr+\bm p\cdot\bm r))\nonumber\\
&&+\frac{\pi\eta^2{\mbox e}^{i\frac{\pi}{4}}}{2\sqrt{2pr}}\Gamma(1/2-i\eta)F(1/2+i\eta,1,\,i(pr+\bm p\cdot\bm r))]\,,\nonumber\\
&&F_B(\bm r,\,\bm p,\,\eta)=-\frac{i}{2}\exp\left(\frac{\pi\eta}{2}-i\bm p\cdot\bm r\right)
[\Gamma(1-i\eta)F(1+i\eta,\,2,\,i(pr+\bm p\cdot\bm r))\nonumber\\
&&+\frac{\pi\eta^2{\mbox e}^{i\frac{\pi}{4}}}{2\sqrt{2pr}}\Gamma(1/2-i\eta)F(3/2+i\eta,\,2,\,i(pr+\bm p\cdot\bm r))]\,,\nonumber\\
&&{\tilde F}_B(\bm r,\,\bm p,\,\eta)=\frac{1}{2}\exp\left(\frac{\pi\eta}{2}-i\bm p\cdot\bm r\right)
[\Gamma(2-i\eta)F(i\eta,\,2,\,i(pr+\bm p\cdot\bm r))\nonumber\\
&&+\frac{\pi\eta^2{\mbox e}^{i\frac{\pi}{4}}}{2\sqrt{2pr}}\Gamma(3/2-i\eta)F(1/2+i\eta,2,\,i(pr+\bm p\cdot\bm r))]\,.
\end{eqnarray}
Here $\Gamma(x)$ is the Euler gamma function, and $F(\alpha,\beta,x)$   is the confluent hypergeometric  function.
Then we use the approach of Ref. \cite{OM59} based on the integral taken in Ref. \cite{N54},
\begin{eqnarray}\label{Nord}
&&\int  \frac{d\bm r}{r}\,\exp\left[-i\bm Q\cdot\bm r-i\frac{m^2\omega}{2\varepsilon_p\varepsilon_q}\,\lambda r\right]
 F(-ia_1,\,1,\, i(qr+\bm q\cdot\bm r))\, F(ia_2,\,1,\, i(pr+\bm p\cdot\bm r))
\nonumber\\
&&=\frac{4\pi}{Q^2}\left(\frac{m^2\omega (1+\xi_p\lambda)}{\varepsilon_p\xi_p Q^2}\right)^{ia_1}
\left(\frac{m^2\omega (1+\xi_q\lambda)}{\varepsilon_q\xi_q Q^2}\right)^{-ia_2}
F(-ia_1,ia_2,1,z)\,,\nonumber\\
&&\quad z=1-\frac{Q^2\xi_p\xi_q(1+\lambda)}{m^2(1+\xi_p\lambda)(1+\xi_q\lambda)}\,,\quad
\xi_p=\frac{1}{1+\delta_p^2}\,,\quad \xi_q=\frac{1}{1+\delta_q^2}\,,
\nonumber\\
&&\bm\delta_p=\frac{\varepsilon_p\bm\theta_p}{m}\,,\quad \bm\delta_q=\frac{\varepsilon_q\bm\theta_q}{m}\,,\quad
\bm Q=\bm p+\bm q-\bm k\,.
\end{eqnarray}
Here we assume that $|\lambda|\sim 1$. We write $g=g_0+\delta g$ and $\bm g^{(\pm)}=\bm g^{(\pm)}_0+\delta\bm g^{(\pm)}$ where            the leading terms are
\begin{eqnarray}\label{gpmI}
&&g_0=N[(\xi_q-\xi_p)\,i\eta {\cal F}+(1-\xi_p-\xi_q)\,(1-u){\cal F}' ]\,,\nonumber\\
&&\bm g^{(\pm)}_0=N\,\frac{(\varepsilon_p\mp\varepsilon_q)}{\omega}[(\xi_p\bm\delta_p+\xi_q\bm\delta_q)\,i\eta {\cal F}
+(\xi_p\bm\delta_p-\xi_q\bm\delta_q)\,(1-u){\cal F}' ]\,,\nonumber\\
&&
N=-i\frac{2\pi}{mQ^2}\left(\frac{\varepsilon_q\xi_q}{\varepsilon_p\xi_p}\right)^{i\eta}|\Gamma(1-i\eta)|^2\,,\nonumber\\
&&{\cal F}=F(-i\eta,\,i\eta,\,1,\, u)\,,\quad
{\cal F}'=\frac{\partial{\cal F}}{\partial u}\,,\quad u=1-\frac{\bm Q^2_\perp}{m^2}\xi_p\xi_q\,.
\end{eqnarray}
Here $\eta=Z\alpha$ and $F(a,b,c,x)$ is the hypergeometric  function.

The first quasiclassical corrections, $\delta g$ and $\delta\bm g^{(\pm)}$, are given by the integrals
\begin{eqnarray}\label{deltaIgpmr}
&&\delta g=\frac{\pi\eta^2{\mbox e}^{i\frac{\pi}{4}}}{8\sqrt{2}}\,\int\frac{d\bm r}{\sqrt{r}}\exp(-i\bm Q\cdot\bm r)
\Bigg\{\frac{1}{\sqrt{\varepsilon_q}}\Gamma(1-i\eta)\Gamma(1/2+i\eta)\nonumber\\
&&\times F(i\eta,\,1,\,i(pr+\bm p\cdot\bm r))F(1/2-i\eta,\,1,\,i(qr+\bm q\cdot\bm r))+(\bm p\longleftrightarrow \bm q\,,\,
\eta\longrightarrow -\eta)\Bigg\}\,,\nonumber\\
&&\delta\bm g^{(\pm)}=\frac{\pi\eta^2{\mbox e}^{i\frac{\pi}{4}}(\varepsilon_p\mp \varepsilon_q)}{8\sqrt{2\varepsilon_p\varepsilon_q}}\,\int\frac{d\bm r}{\sqrt{r}}\exp(-i\bm Q\cdot\bm r)
\Bigg\{\frac{1}{\sqrt{\varepsilon_p}}\Gamma(1-i\eta)\Gamma(1/2+i\eta)\nonumber\\
&&\times F(i\eta,\,1,\,i(pr+\bm p\cdot\bm r))\Bigg[(\bm n_\perp+\bm\theta_q)(1/2+i\eta) F(1/2-i\eta,\,2,\,i(qr+\bm q\cdot\bm r))
\nonumber\\
&&-\bm n_\perp\, F(1/2-i\eta,\,1,\,i(qr+\bm q\cdot\bm r))\Bigg]-(\bm p\leftrightarrow \bm q\,,\, \eta\rightarrow -\eta)\Bigg\}\,.
\end{eqnarray}

In order to take the integral over $\bm r$,  we use the parameterization
\begin{equation}\label{par}
\frac{1}{\sqrt{r}}=\frac{\mbox e^{i\pi/4}}{\sqrt{\pi}}\int\limits_0^\infty
\frac{d\lambda}{\sqrt{\lambda}}\mbox e^{-i\lambda r}\,.
\end{equation}
Then, we  obtain
\begin{eqnarray}\label{deltag}
&&\delta g=\frac{\pi^{3/2}\eta^2}{2mQ}\left(\frac{\varepsilon_q\xi_q}
{\varepsilon_p\xi_p}\right)^{i\eta}\int\limits_0^\infty\frac{d\lambda}{\sqrt{\lambda}}
\left(\frac{1+\xi_p\lambda}{1+\xi_q\lambda}\right)^{i\eta}\nonumber\\
&&\times\Bigg\{\frac{\sqrt{\xi_p}\Gamma(1-i\eta)\Gamma(1/2+i\eta)}{\varepsilon_q\sqrt{1+\xi_p\lambda}}
\bigg[\bigg((1/2-i\eta)\frac{\xi_p}{1+\xi_p\lambda} +i\eta \frac{\xi_q}{1+\xi_q\lambda}\bigg)
 {\cal G}\nonumber\\
 &&+\bigg(\frac{1}{1+\lambda}-\frac{\xi_p}{1+\xi_p\lambda}-\frac{\xi_q\bm}{1+\xi_q\lambda}\Bigg)\,
 (1-z){\cal G}'\bigg]+
 (\bm p\leftrightarrow \bm q\,,\, \eta\rightarrow -\eta)\Bigg\}\,,\nonumber\\
&&\delta\bm g^{(\pm)}=\frac{\pi^{3/2}\eta^2(\varepsilon_p\mp\varepsilon_q)}{2mQ\omega}\left(\frac{\varepsilon_q\xi_q}
{\varepsilon_p\xi_p}\right)^{i\eta}\int\limits_0^\infty\frac{d\lambda}{\sqrt{\lambda}}
\left(\frac{1+\xi_p\lambda}{1+\xi_q\lambda}\right)^{i\eta}\nonumber\\
&&\times\Bigg\{\frac{\sqrt{\xi_p}\Gamma(1-i\eta)\Gamma(1/2+i\eta)}{\varepsilon_q\sqrt{1+\xi_p\lambda}}
\bigg[\bigg(-(1/2-i\eta)\frac{\xi_p\bm\delta_p}{1+\xi_p\lambda} +i\eta \frac{\xi_q\bm\delta_q}{1+\xi_q\lambda}\bigg)
 {\cal G}\nonumber\\
 &&+\bigg(\frac{\xi_p\bm\delta_p}{1+\xi_p\lambda}-\frac{\xi_q\bm\delta_q}{1+\xi_q\lambda}\Bigg)\,
 (1-z){\cal G}'\bigg]-
 (\bm p\leftrightarrow \bm q\,,\, \eta\rightarrow -\eta)\Bigg\}\,,\nonumber\\
&&{\cal G}=F(1/2-i\eta,\,i\eta,\,1,\, z)\,,\quad
{\cal G}'=\frac{\partial{\cal G}}{\partial z}\,.
\end{eqnarray}
Here $z$ is defined in Eq. (\ref{Nord}).

\section{Calculation of the photoproduction cross section}

Using the matrix element obtained it is easy to write the cross section with all polarizations taken into account.
For the cross section summed  over the polarization of the electron and positron, it is necessary to calculate
\begin{equation}\label{M2}
\sum_{\lambda_{1,2}}|M_{\lambda_1\lambda_2}|^{2}=2[|a_0+a_1|^2+|\bm b_0+\bm b_1|^2]=
2[|a_0|^2+|\bm b_0|^2+2\mbox {Re}(a_0a_1^*+\bm b_0\cdot\bm b_1^*)]\,,
\end{equation}
where we neglect the terms $|a_1|^2$ and $|\bm b_1|^2$. It follows from Eq. (\ref{aabb}) that
$$\mbox {Re}(a_0a_1^*+\bm b_0\cdot\bm b_1^*)=0$$
 for any photon polarization. Thus, the terms with $a_1$ and $\bm b_1$ do not contribute to the next-to-leading correction to the cross section summed over the electron and positron polarizations.
For simplicity, we  restrict ourselves to the case of unpolarized photon.
From Eqs. (\ref{aabb}), (\ref{M2}), (\ref{gpmI}), and (\ref{deltag}), we have
\begin{equation}\label{sec1}
d\sigma=\frac{\alpha m^4\, d\varepsilon_p\,d\bm\delta_p\,d\bm\delta_q}{2\pi^4 \omega^3}[(\varepsilon_p^2+\varepsilon_q^2)
|\bm g^{(-)}|^2+\omega^2|g|^2],
\end{equation}
where we used the relation $\bm g^{(+)}=(\varepsilon_p-\varepsilon_q)\bm g^{(-)}/\omega$, see Eqs. (\ref{gpmI}) and (\ref{deltag}).
We write $d\sigma=d\sigma_s+d\sigma_a$, where the leading term is
\begin{eqnarray}\label{secs}
&&d\sigma_s=\frac{2\alpha m^2|\Gamma(1-i\eta)|^4\, d\varepsilon_p\,d\bm\delta_p\,d\bm\delta_q}{\pi^2 \omega^3Q^4}\nonumber\\
&&\times\bigg\{[(1-u)(\varepsilon_p^2+\varepsilon_q^2)+2\varepsilon_p\varepsilon_q(\xi_p-\xi_q)^2]\eta^2{\cal F}^2\nonumber\\
&&+[u(\varepsilon_p^2+\varepsilon_q^2)+2\varepsilon_p\varepsilon_q(1-\xi_p-\xi_q)^2](1-u)^2{{\cal F}'}^2\bigg\}\,.
\end{eqnarray}
Here $u$ and $\cal F$ are defined in Eq.~(\ref{gpmI}). The leading term is symmetric with respect to replacement $\bm p\leftrightarrow \bm q$. The correction $d\sigma_a$ has the form
\begin{eqnarray}\label{secnl}
&&d\sigma_a=-\frac{\alpha m^2\eta^2|\Gamma(1-i\eta)|^2\, d\varepsilon_p\,d\bm\delta_p\,d\bm\delta_q}{2\pi^{3/2} \omega^3Q^3}\,{\mbox{ Im}}\Bigg\{\,\int\limits_0^\infty\frac{d\lambda}{\sqrt{\lambda}}
\left(\frac{1+\xi_p\lambda}{1+\xi_q\lambda}\right)^{i\eta}\nonumber\\
&&\times\frac{\sqrt{\xi_p}\Gamma(1-i\eta)\Gamma(1/2+i\eta)}{\varepsilon_q\sqrt{1+\xi_p\lambda}}{\cal M}+
(\bm p\leftrightarrow \bm q\,,\, \eta\rightarrow -\eta)\Bigg\}\,,\nonumber\\
&&{\cal M}=[(\xi_p-\xi_q)\,i\eta {\cal F}+(1-\xi_p-\xi_q)\,(1-u){\cal F}' ]\nonumber\\
&&\times[4\varepsilon_p\varepsilon_q(\xi_pf_1+\xi_qf_2+f_3)+(\varepsilon_p^2+\varepsilon_q^2)
(f_1+f_2+2f_3)]\nonumber\\
&&+(\varepsilon_p^2+\varepsilon_q^2)(1-u)[(f_1-f_2)i\eta{\cal F}-u(f_1+f_2){\cal F}']\,,\nonumber\\
&&f_1=\frac{(1/2-i\eta){\cal G}-(1-z){\cal G}'}{1+\xi_p\lambda}\,,\quad
 f_2=\frac{i\eta{\cal G}-(1-z){\cal G}'}{1+\xi_q\lambda}\,,\nonumber\\
 &&f_3=\frac{(1-z){\cal G}'}{1+\lambda}\,.
\end{eqnarray}
Here $z$ and $\cal G$ are defined in Eq. (\ref{Nord}). As should be, the correction $d\sigma_a$ is invariant under the replacement $\bm p\leftrightarrow \bm q\,,\, \eta\rightarrow -\eta$. Since $i$ enters this expression only in the combination $i\eta$, it is evident that the correction $d\sigma_a$ is antisymmetric with respect to replacement $\eta\to -\eta$, as well as with respect to replacement $\bm p\leftrightarrow \bm q$.

\section{Special cases}

If $\eta\ll  1$, the leading term $d\sigma_s$ has the form
\begin{eqnarray}\label{secsB}
&&d\sigma_s=\frac{2\alpha m^2\eta^2\, d\varepsilon_p\,d\bm\delta_p\,d\bm\delta_q}{\pi^2 \omega^3Q^4}\nonumber\\
&&\times\left[\frac{Q^2}{m^2}\xi_p\xi_q(\varepsilon_p^2+\varepsilon_q^2)+2\varepsilon_p\varepsilon_q(\xi_p-\xi_q)^2\right]\,.
\end{eqnarray}
 The correction Eq. (\ref{secnl}) at $\eta\ll  1$ reads
\begin{eqnarray}\label{secaB}
&&d\sigma_a=-\,\frac{\alpha m^2\eta^3 d\varepsilon_p\,d\bm\delta_p\,d\bm\delta_q}{2\pi\omega^3Q^3}\,
\bigg\{(\xi_p-\xi_q)\bigg[4(\varepsilon_p\xi_p+\varepsilon_q\xi_q)+
\frac{\omega(\varepsilon_p^2+\varepsilon_q^2)}{\varepsilon_p\varepsilon_q}\bigg]
\nonumber\\
&&+(\varepsilon_p-\varepsilon_q)\frac{(\varepsilon_p^2+\varepsilon_q^2)}{\varepsilon_p\varepsilon_q}\,\frac{Q^2}{m^2}
\xi_p\xi_q\bigg\}\,.
\end{eqnarray}
In the limit $\delta_p\gg 1$  and $\delta_q\gg 1$ this formula reduces to
\begin{eqnarray}\label{secaBas}
d\sigma_a&=&-\,\frac{\alpha \eta^3(\varepsilon_p^2+\varepsilon_q^2) d\varepsilon_p\,d\bm\delta_p\,d\bm\delta_q}{2\pi\varepsilon_p\varepsilon_q\delta_p^2\delta_q^2\omega^3Q^3}\,
[m^2(\delta_q^2-\delta_p^2)\omega
+(\varepsilon_p-\varepsilon_q)Q^2]\nonumber\\
&=&-\,\frac{\alpha \eta^3(\varepsilon_p^2+\varepsilon_q^2) d\varepsilon_p\,d\bm\delta_p\,d\bm\delta_q}{\pi\delta_p^2\delta_q^2\omega^3Q^3}(\bm\theta_q-\bm\theta_p)\cdot{\bm Q}\,
\,.
\end{eqnarray}
The correction $d\sigma_a$ at $\eta\ll 1$ ,  $\delta_p\gg 1$ , and $\delta_q\gg 1$ was also investigated in Ref. \cite{BrGil68} in scalar electrodynamics. Our result (\ref{secaBas}), obtained for fermions, differs from the result of Ref. \cite{BrGil68} for scalar particles by the factor $(\varepsilon_p^2+\varepsilon_q^2)/(\varepsilon_p\varepsilon_q)$. This factor is equal to $2$ for  $|\varepsilon_p-\varepsilon_q|\ll \omega$ in accordance with the statement of Ref. \cite{BrGil68}.

From the experimental point of view, it may be interesting to consider the case $\delta_p\gg\delta_q\gg1$ or $\delta_q\gg\delta_p\gg1$ at $\eta\lesssim 1$. Then the leading symmetric term is
\begin{eqnarray}\label{secsLarges}
&&d\sigma_s=\frac{2\alpha \eta^2\xi_p\xi_q\,(\varepsilon_p^2+\varepsilon_q^2)\, d\varepsilon_p\,d\bm\delta_p\,d\bm\delta_q}{\pi^2 \omega^3Q^2}\,,
\end{eqnarray}
where $Q\approx m|\bm\delta_p+\bm\delta_q|$,  $\xi_p\approx 1/\delta_p^2$, and $\xi_q\approx 1/\delta_q^2$. This term is proportional to $\eta^2$ for any $\eta$. The leading antisymmetric term is
\begin{eqnarray}\label{secsLargea}
&&d\sigma_a=-\frac{\alpha m^2\eta^2(\varepsilon_p\xi_p-\varepsilon_q\xi_q)\,(\varepsilon_p^2+\varepsilon_q^2)\, d\varepsilon_p\,d\bm\delta_p\,d\bm\delta_q}{\pi \varepsilon_p\varepsilon_q\omega^3Q^3}\,\mbox{Re}\,g(\eta)\,,\nonumber\\
&&g(\eta)=\eta\frac{\Gamma(1-i\eta)\Gamma(1/2+i\eta)}{\Gamma(1+i\eta)\Gamma(1/2-i\eta)}\,.
\end{eqnarray}

It is also important to consider the asymtotics of the charge asymmetry in the region $|\bm \delta_p+\bm \delta_q|\ll |\bm \delta_p-\bm \delta_q|$. In this case, the arguments $u$ and $z$ of the hypergeometric functions $\cal F$ and $\cal G$, as well as the factor $[(1+\xi_p\lambda)/(1+\xi_q\lambda)]^{i\eta}$, in Eq. (\ref{secnl}) can be replaced by unity. As a result, we find that $d\sigma_s\propto\eta^2$ and $d\sigma_a\propto\eta^3$ for any $\eta$, and one can use Eqs. (\ref{secsB}) and (\ref{secaB}) for this region.

\begin{figure}[h]
\includegraphics[scale=1.3]{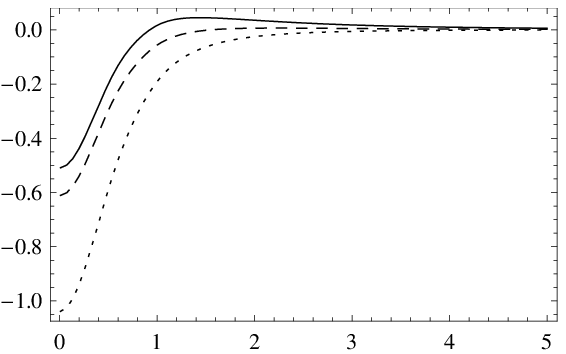}
\begin{picture}(0,0)(0,0)
\put(-110,-15){ $\delta_p$}
 \put(-240,50){\rotatebox{90}{$\frac{m\omega^2}{\alpha}\frac{d\sigma_a}{d\varepsilon_p\,d\bm\delta_p}$}}
 \end{picture}
\caption{The dependence of  $\frac{d\sigma_a}{d\varepsilon_p\,d\bm\delta_p}$ in units $\frac{\alpha}{m\omega^2}$ on $\delta_p$  for a few values of $x=\varepsilon_p/\omega$:  $x=0.25$ (solid curves), $x=0.5$ (dashed curves), and $x=0.75$ (dotted curves); $\eta=0.54$ (tungsten).
}\label{dSadp}
\end{figure}

Integration of Eq. (\ref{secs}) over $\bm\delta_p$  gives  for $d\sigma_s$, Ref. \cite{OM59},
 \begin{eqnarray}\label{secsdeltap}
&&d\sigma_s=\,\frac{4\alpha \eta^2\xi_p^2\,d\bm\delta_p\,d\varepsilon_p}{\pi m^2\omega^3}\left\{(\varepsilon_p^2+\varepsilon_q^2)(L+3/2)+\varepsilon_p\varepsilon_q[1+4\xi_p (1-\xi_p)L\, ]\right\}\,,\nonumber\\
&&L=\ln\left(\frac{2\varepsilon_p\varepsilon_q}{m\omega}\right)-2-\mbox{Re}[\psi(1+i\eta)+C]\,,
\end{eqnarray}
where $C=0.577\ldots$ is the Euler constant. The  integration  over $\bm\delta_p$  gives the well-known result, Ref. \cite{DBM1954},
\begin{eqnarray}\label{secsBspectr}
&&d\sigma_s=\,\frac{4\alpha \eta^2}{m^2\omega^3}\left(\varepsilon_p^2+\varepsilon_q^2+\frac{2}{3}\varepsilon_p\varepsilon_q\right)
\left(L+\frac{3}{2}\right)\,d\varepsilon_p\,.
\end{eqnarray}
We have performed numerical integration of Eq. (\ref{secnl}) over  $\bm\delta_q$. Fig. \ref{dSadp} shows the result of this integration, $\frac{d\sigma_a}{d\varepsilon_p\,d\bm\delta_p}$ in units $\frac{\alpha}{m\omega^2}$ as a function of $\delta_p$  for a few values of $x=\varepsilon_p/\omega$.

The cross section $d\sigma_a$ integrated over both $\bm\delta_p$ and $\bm\delta_q$  was obtained in  our paper  Ref. \cite{LMS2004} and has the form
\begin{eqnarray}\label{secaspectr}
&&d\sigma_a=-\,\frac{\pi^3\alpha \eta^2(\varepsilon_p-\varepsilon_q)
(2\omega^2-3\varepsilon_p\varepsilon_q)\,d\varepsilon_p }{4m\omega^3\varepsilon_p\varepsilon_q}\,\mbox{Re}\,g(\eta)\,.
\end{eqnarray}
The result of numerical integration of Eq. (\ref{secnl}) over $\bm\delta_p$ and $\bm\delta_q$ is in agreement with the above result.

\section{Compton-type contribution}

In this Section we estimate the contribution of the  Compton-type  amplitude to the photoproduction cross section.
\begin{figure}[h]
\includegraphics[width=10cm]{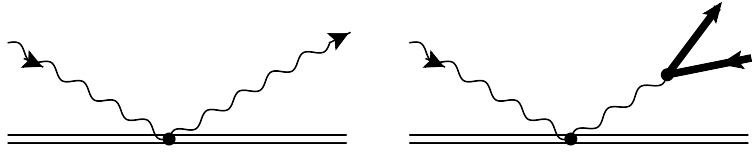}
\caption{Real Compton scattering diagram (left) and
Compton-type diagram for pair production by a photon in a strong Coulomb field (right). Thick lines correspond to the positive- and negative-energy solutions of the Dirac equation in the Coulomb field. Double line denotes nucleus.}\label{fig:dia2}
\end{figure}
Since this amplitude is small in comparison with the leading amplitude found above, it is necessary to take into account only the interference between the  Compton-type  amplitude and  the   leading amplitude. The leading amplitude is   enhanced at small angles $\theta_p$ and $\theta_q$ of the final particles. Therefore, it is sufficient to calculate the Compton-type  amplitude also at $\theta_p,\,\theta_q\ll1$. For real initial and final photons with $\omega\ll m_A$ ($m_A$ is the nuclear mass), the nuclear  Compton scattering amplitude, corresponding to the left diagram in Fig.\ref{fig:dia2}, in the forward direction reads
\begin{eqnarray}\label{Comptonampl}
M_{C}=T(\omega)\,\bm e_1\cdot\bm e_2^*\,,
\end{eqnarray}
where $T(\omega)$ is the function measured for many nuclei, see Review \cite{HLMS2000},  $\bm e_1$ and $\bm e_2$ are the photon  polarization vectors of the initial and final photons, respectively. The function $T(\omega)$ satisfies the relations
\begin{eqnarray}\label{Aomega}
&&T(0)=-\frac{Z^2e^2}{m_A}\,,\quad \mbox{Im}\,T(\omega)=\frac{\omega}{4\pi}\sigma_{\gamma N}(\omega)\,,\nonumber\\
&&
\mbox{Re}\,[T(\omega)-T(0)]=\frac{\omega^2}{2\pi^2}\,{\cal P}\!\!\int_0^\infty\frac{\sigma_{\gamma N}(\omega)}{\omega'^2-\omega^2}
\,d\omega'\,,
\end{eqnarray}
where $\sigma_{\gamma N}(\omega)$ is the nuclear photoabsorption cross section and ${\cal P}$ denotes  the integration in the principal value sense.  Below pion photoproduction threshold, the cross section $\sigma_{\gamma N}(\omega)$ is conventionally written as a superposition of Lorentzian lines
\begin{eqnarray}\label{siggn}
\sigma_{\gamma N}(\omega)=\sum_n\sigma_n\frac{(\omega\Gamma_n)^2}{(E_n^2-\omega^2)^2+(\omega\Gamma_n)^2}\,,
\end{eqnarray}
where the parameters $\sigma_n$, $E_n$, and $\Gamma_n$ are extracted from the experiment. The corresponding function $T(\omega)$
has the form
\begin{eqnarray}\label{AomegaL}
T(\omega)=-\frac{Z^2e^2}{m_A}+\frac{\omega^2}{4\pi}\sum_n
\frac{\sigma_n\Gamma_n}{E_n^2-\omega^2-i\omega\Gamma_n}\,.
\end{eqnarray}
Below pion threshold, but above the resonance region ($\omega\gg E_n$), the function $T(\omega)$ has the form
\begin{eqnarray}\label{AomegaL1}
T(\infty)=-\frac{Z^2e^2}{m_A}-\frac{1}{4\pi}\sum_n\sigma_n\Gamma_n
\,.
\end{eqnarray}
The last term in this asymptotics is equal to $(1+\varkappa)NZe^2/m_A$, where $\varkappa$ is the so called enhancement factor, see Ref.~\cite{HLMS2000}, and
$N$ is the number of neutrons, and we obtain
\begin{equation}
T(\infty)=-\frac{Ze^2}{m_p}(1+\frac{N}{A}\varkappa)\,,
\end{equation}
where $m_p$ is the proton mass, $A=Z+N$. For heavy nuclei $\varkappa\sim 0.3\div 0.4$, Ref.~\cite{Schu93}, so that
$$T(\infty)/T(0)\sim 3\,.$$

Using the function  $T(\omega)$,  we write the additional  contribution  to the photoproduction amplitude, corresponding to the right diagram in Fig.\ref{fig:dia2}, as follows
\begin{eqnarray}\label{Thomsonampl}
&&{\tilde M}_{\lambda_1\lambda_2}\,=\,-T(\omega)\int \frac{d\bm \kappa}{(2\pi)^3}
 \frac{4\pi}{\omega^2-\kappa^2+i\,0}\nonumber\\
&&\times \left(\bm e-\frac{(\bm\kappa\cdot\bm e)}{\omega^2}\bm\kappa\right)\cdot\,\int d\bm r \,\bar u_{\lambda_1\bm p }^{(out)}(\bm r )\,\bm\gamma
\,v _{\lambda_2\bm q}^{(in)}(\bm r )\exp{(i\bm \kappa\cdot\bm r )}\,\,.
\end{eqnarray}
We assume that $\theta_p\ll1$ and $\theta_q\ll1$. Taking the integral over $\bm\kappa$ we obtain
\begin{eqnarray}\label{Thomsonampl1}
&&{\tilde M}_{\lambda_1\lambda_2}\,=\,T(\omega)
\,\int \frac{d\bm r}{r} \,\bar u_{\lambda_1\bm p }^{(out)}(\bm r )\left[\bm e-(\bm n\cdot\bm e)\bm n\right] \cdot\bm\gamma
\,v _{\lambda_2\bm q}^{(in)}(\bm r )\exp{(i\omega  r )}\,,
\end{eqnarray}
where $\bm n=\bm r/r$. The main contribution to the integral over $\bm r$ is given by the region $\bm p\cdot\bm r\sim pr\sim \omega^2/m^2$ and $\bm q\cdot\bm r\sim qr\sim \omega^2/m^2$. In this case
\begin{eqnarray}\label{fun123as1}
&&F_A(\bm r,\,\bm p,\,\eta)={\tilde F}_B(\bm r,\,\bm p,\,\eta)=\frac{\exp(-i\bm p\cdot\bm r)}{2(pr+\bm p\cdot\bm r)^{i\eta}}
\,,\nonumber\\
&&F_B(\bm r,\,\bm p,\,\eta)=\frac{\exp(-i\bm p\cdot\bm r)}{2(pr+\bm p\cdot\bm r)^{i\eta+1}}\,,
\end{eqnarray}
Using this asymptotics and taking the integral over $\bm r$ we arrive at
the Compton-type correction to the photoproduction amplitude
\begin{eqnarray}\label{Thomsonampl2}
&&{\tilde M}_{\lambda_1\lambda_2}\,=\phi_{\lambda_1}^+(\tilde{a}+\bm\sigma\cdot\tilde{\bm b})\chi_{\lambda_2}\,,\nonumber\\
&&\tilde{a}=i \tilde{N}\,[\bm\nu\times\bm e]\cdot\bm\vartheta\,,\quad \tilde{\bm b}=\tilde{N}\frac{(\varepsilon_p-\varepsilon_q)}{\omega}
(\bm e\cdot\bm\vartheta)\,\bm\nu\,,\nonumber\\
&&\tilde{N}=\,\frac{2\pi\,T(\omega)}{m\omega}\left(\frac{\varepsilon_q }{\varepsilon_p}\right)^{i\eta}
\frac{1}{1+\vartheta^2}\,,\nonumber\\
&&\bm\vartheta=\frac{\varepsilon_p\varepsilon_q}{m\omega}(\bm\theta_p-\bm\theta_q)=
\frac{\varepsilon_q\bm\delta_p-\varepsilon_p\bm\delta_q}{\omega}\,.
\end{eqnarray}
The corresponding correction to the cross section has the form
\begin{eqnarray}\label{secCompton}
&&d\tilde\sigma\,=
\frac{4\alpha m^4\,d\varepsilon_p\,d\bm\delta_p\,d\bm\delta_q}{(2\pi)^4\omega}\mbox{Re}(a_0 \tilde{a}^*+\bm b_0\cdot\tilde{\bm b}^*)\nonumber\\
&&=\frac{8\alpha m^4(\varepsilon_p^2+\varepsilon_q^2)\,d\varepsilon_p\,d\bm\delta_p\,d\bm\delta_q}{(2\pi)^4\omega^3}
\mbox{Re}[(\bm g^{(-)}_0\cdot\bm\vartheta) \tilde{N}^*]\,.
\end{eqnarray}
This correction contains both symmetric and antisymmetric parts with respect to replacement $\eta\rightarrow -\eta$.
The symmetric part is proportional to $\mbox{Im}\,T(\omega)$ and the antisymmetric part is proportional to $\mbox{Re}\,T(\omega)$
\begin{eqnarray}\label{secComptona}
&&d\tilde\sigma_a\,=
\frac{2\alpha m^2|\Gamma(1-i\eta)|^2(\varepsilon_p^2+\varepsilon_q ^2)\,d\varepsilon_p\,d\bm\delta_p\,d\bm\delta_q}{\pi^2\omega^4(1+\vartheta^2)Q^2}\,\mbox{Re}\,T(\omega) \nonumber\\
&&\times\bigg(\bm\vartheta\cdot\left[\cos\mu\,(\xi_p\bm\delta_p+\xi_q\bm\delta_q)\,\eta {\cal F}
+\sin\mu\,(\xi_p\bm\delta_p-\xi_q\bm\delta_q)\,(1-u){\cal F}' \right]\bigg)\,,\nonumber\\
&&\mu =\eta\ln\left(\frac{\xi_q}{\xi_p}\right)\,.
\end{eqnarray}


\section{ Discussion}

\begin{figure}[h]
\includegraphics[scale=1.4]{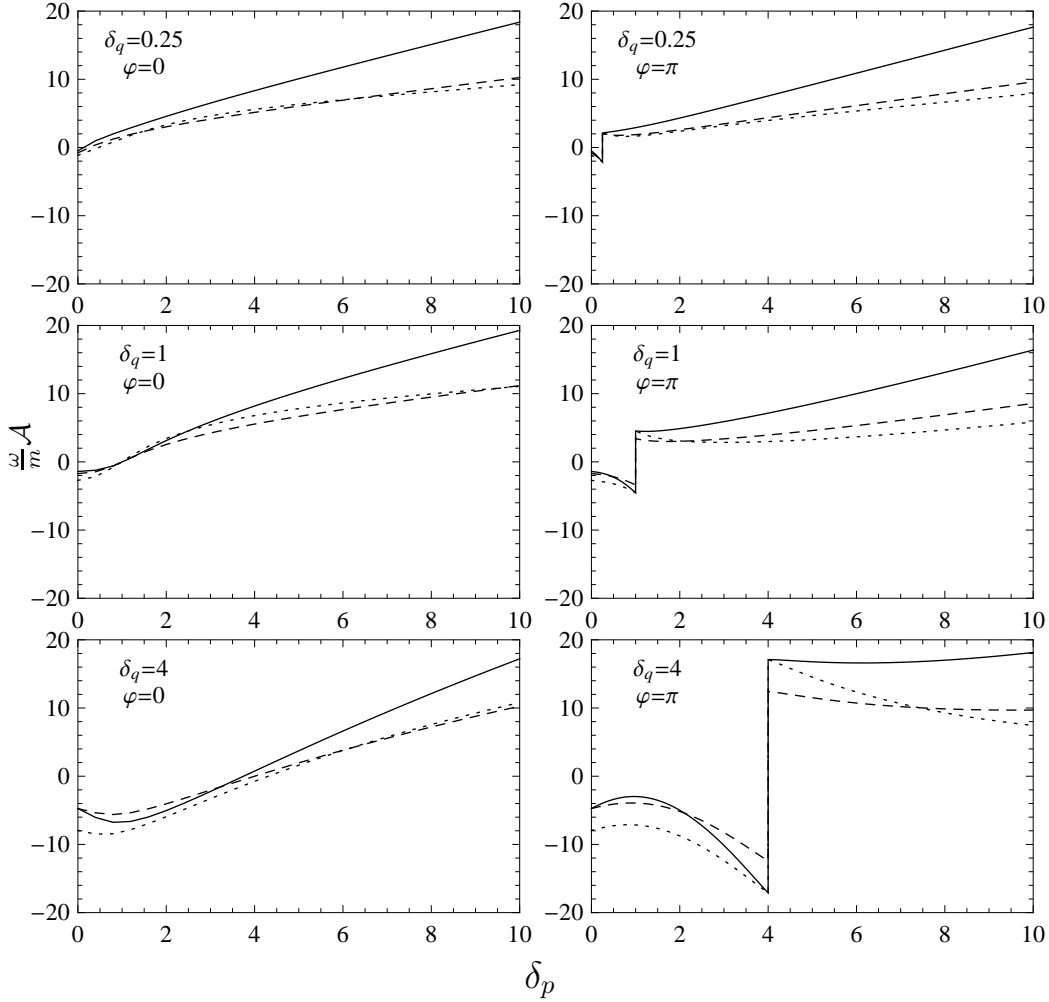}
\begin{picture}(0,0)(0,0)
\put(-205,-15){\large $\delta_p$}
 \put(-400,180){\rotatebox{90}{$\frac{\omega}{m}{\cal A}$}}
 \end{picture}
\caption{The dependence of  ${\cal A}$ on $\delta_p$ in units $m/\omega$ for a few values of $\delta_q$, $\varphi$, and $x=\varepsilon_p/\omega$: $x=0.25$ (solid curves), $x=0.5$ (dashed curves), and $x=0.75$ (dotted curves), $\eta=0.54$ (tungsten).
}\label{SigmaDD}
\end{figure}

In quantum electrodynamics, an electron differs from a positron only by its charge. Therefore, the cross section of $e^+e^-$ pair photoproduction satisfies the relation
$$d\sigma(\bm p,\, \bm q,\, \eta)=d\sigma(\bm q,\, \bm p,\, -\eta).$$
We define the charge asymmetry ${\cal A}$ as
\begin{equation}\label{asR}
{\cal A}=\frac{d\sigma(\bm p,\, \bm q,\, \eta)-d\sigma(\bm p,\, \bm q,\, -\eta)}
{d\sigma(\bm p,\, \bm q,\, \eta)+d\sigma(\bm p,\, \bm q,\, -\eta)}=\frac{d\sigma(\bm p,\, \bm q,\, \eta)-d\sigma(\bm q,\, \bm p,\, \eta)}
{d\sigma(\bm p,\, \bm q,\, \eta)+d\sigma(\bm q,\, \bm p,\, \eta)} \,.
\end{equation}
Let us first neglect $d\tilde\sigma_a$ and  calculate  $\cal R$ as a ratio of $d\sigma_a$ in Eq. (\ref{secnl}) and $d\sigma_s$ in Eq. (\ref{secs}). Outside the very narrow region $Q_{\perp}\lesssim |Q_{\parallel}|=|\bm\nu\cdot\bm Q|\sim m^2/\omega$, we can replace $Q^2\to \bm Q^2_{\perp}=m^2(\bm\delta_p+\bm\delta_q)^2$. Then, at fixed $\bm \delta_p$, $\bm \delta_q$, and $x=\varepsilon_p/\omega$ the asymmetry $\cal A$ scales as $m/\omega$, as can be seen from Eqs. (\ref{secnl}) and (\ref{secs}). Figure \ref{SigmaDD} shows the dependence of ${\cal A}$ on $\delta_p$ in units $m/\omega$ for tungsten ( $\eta=0.54$) for a few values of $\delta_q$ and $\varphi$, where $\varphi$ is the angle between vectors $\bm\delta_p$ and
$\bm\delta_q$. It is seen that the charge asymmetry may be rather large ( ${\cal A}\sim 20\div 30\%$ for $\omega/m=50$). The asymmetry is large when $\delta_p$ and/or $\delta_q$ are much larger than unity. Note that this statement is also valid in the region $|\bm \delta_p+\bm \delta_q|\ll |\bm \delta_p-\bm \delta_q|$ (but $\delta_p\gg1$ and $\delta_q\gg 1$).

\begin{figure}[h]
\includegraphics[scale=1.8]{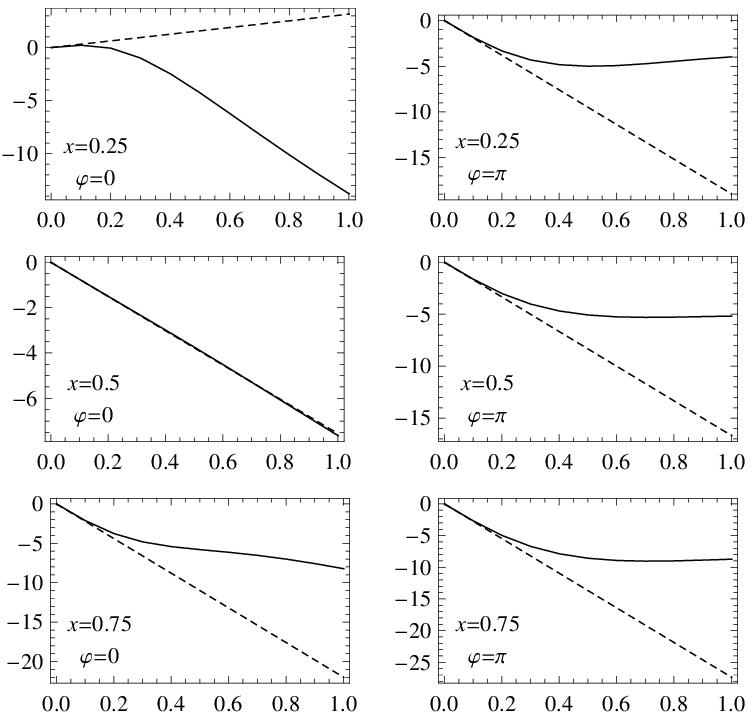}
\begin{picture}(0,0)(0,0)
\put(-200,-15){\large $\eta$}
 \put(-410,180){\rotatebox{90}{$\frac{\omega}{m}{\cal A}$}}
 \end{picture}
\caption{The dependence of  ${\cal A}$ on $\eta=Z\alpha$ in units $m/\omega$ for $\delta_p=2$, $\delta_q=4$ and a few values of  $x$ and $\varphi$. Solid curve represents the exact in $\eta$ result,  dashed curve is obtained in the leading in $\eta$ approximation (linear in $\eta$).
}\label{Zadependence}
\end{figure}

\begin{figure}[h]
\includegraphics[scale=1.3]{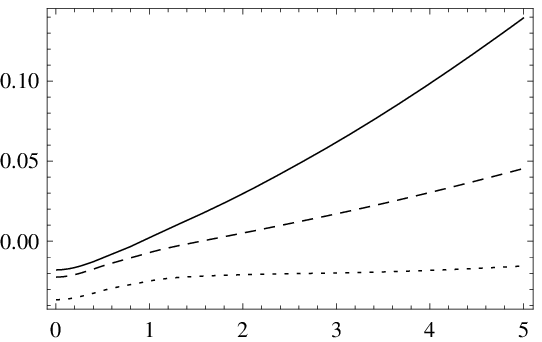}
\begin{picture}(0,0)(0,0)
\put(-110,-15){ $\delta_p$}
 \put(-220,60){\rotatebox{90}{${\cal A}_1$}}
 \end{picture}
\caption{The dependence of  ${\cal A}_1$ on $\delta_p$  for $x=0.25$ (solid curves), $x=0.5$ (dashed curves), and $x=0.75$ (dotted curves); $\eta=0.54$ (tungsten), $\omega/m=50$.
}\label{SigmaDp}
\end{figure}

It is interesting to understand the importance of high-order in $\eta$ terms in the charge asymmetry. Figure \ref{Zadependence}
 shows the dependence of  ${\cal A}$ on $\eta=Z\alpha$ in units $m/\omega$ for $\delta_p=2$, $\delta_q=4$ and a few values of  $x=\varepsilon_p/\omega$ and $\varphi$. The dashed curve in this figure is obtained in the leading in $\eta$ approximation (linear in $\eta$). It is seen that $\eta$ dependence is very strong even for intermediate values of $\eta$.
Let us also introduce  the charge asymmetry ${\cal A}_1$ for the cross section integrated over the angles of one of the particles,
\begin{equation}\label{asR1}
{\cal A}_1=\frac{d\sigma(\bm p,\,\eta)-d\sigma(\bm p, -\eta)}
{d\sigma(\bm p,\,\eta)+d\sigma(\bm p, -\eta)} \,.
\end{equation}
We calculate nominator  in  ${\cal A}_1$ integrating  $d\sigma_a$ in Eq. (\ref{secnl}) over $\bm\delta_q$ and denominator using Eq. (\ref{secsdeltap}). Figure \ref{SigmaDp} shows the dependence of ${\cal A}_1$ on $\delta_p$  for  $\eta=0.54$ (tungsten),     $\omega/m=50$, and a few values of  $x$ . Note that ${\cal A}_1$, in contrast to ${\cal A}$, does not scale as $m/\omega$ due to logarithmic dependence of $d\sigma_s(\bm p,\eta)$ on $\omega$, see Eq. (\ref{secsdeltap}). It is seen that  ${\cal A}_1$ is noticeable though it is  smaller than ${\cal A}$.
The charge asymmetry corresponding to the cross section integrated over the angles of both particles (over $\bm\delta_p$ and
$\bm\delta_q$) is very small (see Ref. \cite{LMS2004}).

Let us discuss now the contribution $\tilde{\cal A}$ to the charge asymmetry,
\begin{equation}\label{asRcompton}
\tilde{\cal A}=\frac{d\tilde\sigma_a(\bm p,\, \bm q,\, \eta)}
{d\sigma_s(\bm p,\, \bm q,\, \eta)} \,,
\end{equation}
where $d\tilde\sigma_a(\bm p,\, \bm q,\, \eta)$ is given by Eq. (\ref{secComptona}) and $d\sigma_s(\bm p,\, \bm q,\, \eta)$
is given by Eq. (\ref{secs}).

In the region $\delta_p\sim \delta_q\sim 1$, we have $d\tilde\sigma_a/d\sigma_a\sim m/(\eta m_p)$ and $d\sigma_a/d\sigma_s\sim \eta m/\omega$. In this region, $d\tilde\sigma_a$ may be  comparable with $d\sigma_a$ only for light nucleus  ($\eta\ll 1$), where asymmetry is very small.

In the region $\delta_p\sim 1$, $\delta_q\gg1$, we have $d\tilde\sigma_a/d\sigma_a\sim m/(\eta m_p)$ and
 $d\sigma_a/d\sigma_s\sim \eta \theta_q$. Again, $d\tilde\sigma_a$ may be  comparable with $d\sigma_a$ only for light nucleus where asymmetry is very small.

\begin{figure}[h]
\setlength{\unitlength}{1.5cm}
\includegraphics[width=5\unitlength]{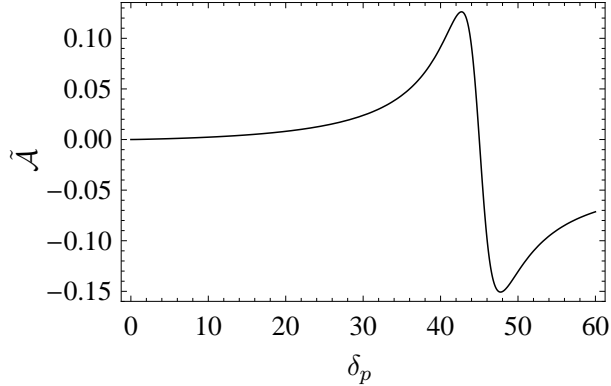}
\begin{picture}(0,0)(0,0)
\put(-2.5,-0.3){ $\delta_p$}
 \put(-5.4,1.5){\rotatebox{90}{$\tilde{\cal A}$}}
 \end{picture}
\caption{Contribution $\tilde {\cal A}$ as a function of $\delta_p$ for $\omega/m=200$, $Z=1$ (proton), $x=\varepsilon_p/\omega=0.6$,  $\delta_q=30$.}\label{fig:tildeA}
\end{figure}

 The only region where  $d\tilde\sigma_a  \gtrsim d\sigma_a$ and $\tilde{\cal A}$ is not too small is $\eta\ll 1$,
 $\delta_p\gg 1$, $\delta_q\gg 1$, but $\vartheta=\frac{\varepsilon_p\varepsilon_q}{m\omega}|\bm\theta_p-\bm\theta_q|\sim1$.
In this region the ratio $d\tilde\sigma_a /d\sigma_a$ is
\begin{eqnarray}\label{tsecaseca}
\frac{d\tilde\sigma_a }{d\sigma_a}=-
\frac{2(1+\frac{N}{A}\varkappa)\varepsilon_p^2\varepsilon_q^2 \theta_p^3}{\pi (1+\vartheta^2)\,\eta\,\omega m^2 m_p}\,,
\end{eqnarray}
and may be larger than unity. Here we took into account that $\theta_p\approx\theta_q\gg m/\omega$ but $|\bm\theta_p-\bm\theta_q|\sim m/\omega$. For the contribution $\tilde{\cal A}$ to the charge asymmetry we have
\begin{eqnarray}\label{rtilde}
\tilde{\cal A}=\frac{d\tilde\sigma_a }{d\sigma_s}=-
\frac{(1+\frac{N}{A}\varkappa)\varepsilon_p\varepsilon_q \theta_p^2\,(\bm\theta_p\cdot\bm\vartheta)}{(1+\vartheta^2) m m_p}\,,
\end{eqnarray}
where $\bm\vartheta=(\bm\theta_p-\bm\theta_q)\varepsilon_p\varepsilon_q/(m\omega)$, so that $\tilde{\cal A}$ may reach about ten percent at large transverse momenta compared to the electron mass, see Fig. \ref{fig:tildeA}.
\section{Conclusion}
We have derived exactly in the parameter $\eta=Z\alpha$ the charge asymmetry ${\cal A}$, Eq. (\ref{asR}), in the process  of $e^+e^-$ photoproduction  in a Coulomb field at photon energy $\omega\gg m$, $\varepsilon_p\gg m$, and $\varepsilon_q\gg m$. This asymmetry is related to the first quasiclassical correction to the differential cross section of the process, Eq. (\ref{secnl}). When $p_\perp$ and/or $q_\perp$ are much larger than the electron mass $m$, the charge asymmetry can be as large as tens percent. The charge asymmetry ${\cal A}_1$, Eq. (\ref{asR1}), in the cross section integrated over the transverse momenta of one of the particles is several times smaller than the asymmetry $\cal A$ in the cross section differential with respect to the transverse momenta of both particles. We have also estimated the contribution $\tilde{\cal A}$, Eqs. (\ref{secComptona}) and (\ref{asRcompton}), to the charge asymmetry of the Compton-type diagram. For $\eta\sim 1$, this contribution is negligible. The only region where $\tilde{\cal A}$ can be important is $\eta\ll 1$ (light nucleus), $\theta_p\approx\theta_q\gg m/\omega$ but $|\bm\theta_p-\bm\theta_q|\sim m/\omega$. Though we have performed our calculation for the pure Coulomb field, our results are also applicable for photoproduction in the electric field of atoms except the very narrow region $Q\lesssim r_{\mathrm{scr}}^{-1}\sim m\alpha Z^{1/3}\ll m$, where $r_{\mathrm{scr}}$ is the atomic screening radius.

Our results clearly demonstrate  that experimental observation of the charge asymmetry in  the process  of $e^+e^-$ photoproduction  in a Coulomb field is a realistic task.

\section*{Acknowledgments}
We thank S.J.~Brodsky and G. Ron for valuable discussions. This work was supported in part by the RFBR Grant
No. 09-02-00024 and the Grant 14.740.11.0082 of Federal special-purpose program ``Scientific and scientific-pedagogical personnel of innovative Russia''.


\begin{thebibliography}{99}
\bibitem{HGO1980} J.~H.Hubbell,
 H.~A. Gimm, and I.~\O{}verb\o{},  J. Phys. Chem. Rev. Data {\bf 9}, 1023 (1980).

\bibitem{Hubbell2000}
J.~H. Hubbell, Rad. Phys. Chem.  {\bf 59}, 113 (2000).

\bibitem{BH1934} H.~A. Bethe and W.~Heitler, Proc. R. Soc. London {\bf A146}, 83 (1934).

\bibitem{Racah1934} G.~Racah, Nuovo Cim. {\bf 11}, 461 (1934).

\bibitem{Overbo1968}
I.~{\O}verb\o{},  K.~J. Mork, and H.~A. Olsen, Phys. Rev. {\bf 175}, 1978 (1968).

\bibitem{SudSharma2006} K. K. Sud and D. K. Sharma, Rad. Phys. and Chem. {\bf 75}, 631 (2006).

\bibitem{BM1954} H.~A. Bethe and L.~C. Maximon, Phys. Rev. {\bf 93}, 768 (1954).

\bibitem{DBM1954} H.~Davies, H.~A. Bethe, and L.~C. Maximon, Phys. Rev. {\bf 93}, 788 (1954).

\bibitem{Overbo1977}
I.~{\O}verb\o{}, Phys.~Lett. {\bf B71}, 412 (1977).

\bibitem{LMS2004} R.~N. Lee, A.~I. Milstein, and V.~M. Strakhovenko, Phys. Rev. A {\bf 69}, 022708 (2004).

\bibitem{LMS2003} R.~N. Lee, A.~I. Milstein, and V.~M. Strakhovenko, {hep-ph/0307388}.

\bibitem{DM10} A.~Di Piazza and A.~I. Milstein, Phys. Rev. A {\bf 82}, 042106 (2010).

\bibitem{BLP82} V.~B.Berestetski, E.~M.Lifshits, L.~P.Pitayevsky, {\it Quantum electrodynamics}
 (Pergamon, Oxford, 1982).
\bibitem{MT2004} A.I. Milstein, I.S. Terekhov, Zh. Eksp. Teor. Fiz. {\bf 125}, 785 (2004) [JETP {\bf 98}, 687 (2004)].
\bibitem{MS82} A.I. Milstein, V.M. Strakhovenko, Phys. Lett.
A \textbf{90} (1982) 447.
\bibitem{OM59} H. Olsen and L.C. Maximon, Phys. Rev.  {\bf 114},  887 (1959).
\bibitem{N54} A. Nordsieck, Phys. Rev.  {\bf 93},  785 (1954).

\bibitem{BrGil68}S.J. Brodsky and J.R. Gillespie, Phys. Rev.  {\bf 173}, 1011 (1968).

\bibitem{HLMS2000} M.T. H\'utt , A.I. L'vov , A.I. Milstein , M. Schumacher,
Phys. Reports. {\bf 323}, 457  (2000).
\bibitem{Schu93} M.~Schumacher, et al., Nucl. Phys. A {\bf 576}, 603 (1994).

\end{thebibliography}
\end{document}